# CRYOPRESERVATION OF PINEAPPLE (*Ananas comosus*) APICES

María Teresa González Arnao[1*], Manfred Márquez Ravelo[2],
Caridad Urra Villavicencio[1], Marcos Martínez Montero[3]
and Florent Engelmann[4].

1: Centro Nacional de Investigaciones Científicas (CNIC), Ave. 25 y 158, Apartado 6990, Cubanacán, Playa, La Habana, Cuba.
2: Universidad de la Habana, Fac. de Biología, Calle 25, e/ I y J, Vedado, La Habana, Cuba.
3: Centro de Bioplantas - Universidad de Ciego de Avila (UNICA), Carretera a Morón 9, CP 69450, Ciego de Avila, Cuba.
4: International Plant Genetic Resources Institute (IPGRI), Via delle Sette Chiese 142, 00145 Rome, Italy.

**Summary:** The encapsulation-dehydration and vitrification techniques were experimented for freezing apices of pineapple *in vitro* plantlets. Positive results were achieved using vitrification only. Optimal conditions included a 2-d preculture of apices on medium supplemented with 0.3M sucrose, loading treatment for 25 min in medium with 0.75M sucrose + 1M glycerol and dehydration with PVS2 vitrification solution at 0°C for 7 h before rapid freezing in liquid nitrogen. This method resulted in 65, 35 and 25% survival with apices of varieties Puerto Rico, Perolera and Smooth Cayenne, respectively. Recovery of cryopreserved apices took place directly, without transitory callus formation.

## INTRODUCTION

Pineapple is a fruit crop of major importance in many tropical countries. Pineapple is vegetatively propagated and crosses between varieties produce botanical seeds. However, these seeds are highly heterozygous and therefore of limited interest for the conservation of specific gene combinations.

Cryopreservation of apices is the most relevant strategy for long-term conservation of vegetatively propagated crops, since true to type, virus-free plants can be regenerated directly from cryopreserved apices. Currently, cryopreservation procedures have been developed for apices of numerous species including many tropical crops (l, 16). The freezing methods employed are principally encapsulation-dehydration and vitrification, which do not require sophisticated equipment for freezing and produce high recovery rates with a wide range of materials.

In this paper, vitrification and encapsulation-dehydration were used to freeze apices of *in vitro* plantlets of pineapple. The most efficient protocol was applied to apices of three different varieties.

## MATERIALS AND METHODS

### Plant material

The experimental material consisted of *in vitro* plantlets of 3 pineapple (*Ananas comosus* (Stickm.) Merr.) varieties (Puerto Rico, Perolera and Smooth Cayenne) from the *in vitro* collection of the Centro de Bioplantas-UNICA. They were obtained by introducing *in vitro* 1 cm-long axillary buds isolated from sterilized field-grown plants and were multiplied on semi-solid MS medium (8) and supplemented with 2 mg.L$^{-1}$ benzylaminopurine (BAP), 0.3mg.L$^{-1}$ naphthalene acetic acid (NAA), 30 g.L$^{-1}$ (0.08M) sucrose and 10 g.L$^{-1}$ agar. The plantlets were maintained at 26°C under an 8 h light/16 h dark photoperiod, with a light intensity of 40 µmol.m$^{-2}$.s$^{-1}$ and were subcultured every 30 d.

For cryopreservation experiments, apices (up to 3 mm in length) were dissected from *in vitro* plantlets 15 d after the last subculture. After dissection, they were either left overnight on standard semi-solid medium or precultured for 2 d on the same medium supplemented with 0.3M sucrose before being used for cryopreservation experiments.

### Cryopreservation

### Encapsulation-dehydration.

Apices which had been submitted to an overnight preculture on standard semi-solid medium were encapsulated in alginate (3%) beads with a concentration of 0.08M sucrose. Apices precultured for 2 d on medium supplemented with 0.3M sucrose were encapsulated in alginate beads with 0.3M sucrose. Apices were then precultured for 24 h in liquid medium with various sucrose concentrations (0.3 to 1.0M) or for extended durations (up to 8 d) in 0.5M sucrose. After pregrowth, the beads were dehydrated at room temperature under the sterile air current of the laminar flow cabinet, down to moisture contents (MC, fresh weight basis) ranging between 35 and 18%, then transferred to sterile 2-ml polypropylene cryotubes and frozen rapidly by direct immersion of the cryotubes in liquid nitrogen. Samples were kept for 1 h at -196°C.

For thawing, the beads were placed in Petri dishes under the air current of the laminar flow cabinet for 2-3 min, then transferred to standard semi-solid medium for regrowth. Apices were cultured for the first week in the dark, then transferred under standard lighting conditions.

*Vitrification procedure*

After dissection, apices were precultured for 2 d on the standard semi-solid medium supplemented with 0.3 M sucrose. They were then loaded at 25°C by placing in a Petri dish on a filter paper imbibed with the cryoprotective solution containing glycerol (1M) and sucrose at various concentrations (0.3, 0.5 and 0.75M). Loading time was 25 min.

After loading, explants were placed in 2-ml cryotubes containing 1 ml of the vitrification solution PVS2 (30% (w/v) glycerol + 15% (w/v) ethylene glycol + 15% (w/v) DMSO + 0.4M sucrose (12) at 25°C or 0°C for different durations. At 25°C, the duration of treatment with PVS2 was between 0 and 7 h. At 0°C, apices were treated for up to 9 h. During treatment with PVS2, the vitrification solution was replaced once with fresh solution. Cryotubes were then plunged rapidly into liquid nitrogen and kept for 1 h at -196°C.

For rewarming, cryotubes were immersed in a water-bath at 40°C for 2-3 min. The PVS2 solution was drained from the cryotubes and samples washed twice with liquid MS medium supplemented with 1.2M sucrose for approximately 30 min. For recovery, apices were transferred to standard semi-solid medium in Petri dishes. They were cultured for the first week in the dark, then transferred under standard lighting conditions.

*Assessment of survival*

Fifteen to 20 apices were employed for each experimental condition. The results presented correspond to the average (±standard deviation) of two independent experiments. Survival was evaluated after 1 month by counting the number of apices which had developed into shoots.

## RESULTS

*Encapsulation-dehydration*

Apices encapsulated in beads containing 0.08M sucrose were extremely sensitive to the sucrose concentration in the pregrowth medium, since concentrations higher than 0.3M led to a drastic drop in survival (Table 1). No or low viability was obtained after desiccation.

With apices encapsulated in beads with 0.3M sucrose, pregrowth with 0.3 and 0.5M resulted in survival rates of 86-90%. Higher sucrose concentrations (0.75 and 1M) led to a drastic decrease in viability.

**Table 1:** Effect of sucrose concentration in the encapsulation and preculture medium on the survival rate (%) of pineapple (variety Puerto Rico) apices after a 24-h pregrowth (P) period and desiccation (D) to 20-24% MC.

| Sucrose concentration in the preculture medium (M) | Encapsulation in beads with 0.08M sucrose | | Encapsulation in beads with 0.3M sucrose | |
|---|---|---|---|---|
| | P | D | P | D |
| 0.3 | 80±8 | 19±6 | 90±4 | 40±9 |
| 0.5 | 53±6 | 9±3 | 86±6 | 38±4 |
| 0.75 | 33±1 | 0 | 60±9 | 10±3 |
| 1.0 | 23±7 | 0 | 31±7 | 0 |

Increasing pregrowth duration in 0.5M sucrose enhanced survival of desiccated apices up to 70%, which was achieved after 5 d of pregrowth. With other treatment durations, viability remained around 50% (Table 2).

**Table 2:** Effect of pregrowth duration in 0.5M sucrose on the survival rate of pineapple (variety Puerto Rico) apices after pregrowth and desiccation to 20-24 MC.

| Pregrowth duration (d) in 0.5M sucrose | Survival (%) | |
|---|---|---|
| | Pregrowth | Desiccation |
| 1 | 86±4 | 38±6 |
| 2 | 70±7 | 35±4 |
| 3 | 66±3 | 40±9 |
| 4 | 83±4 | 50±3 |
| 5 | 90±3 | 70±7 |
| 6 | 88±6 | 40±6 |
| 7 | 90±7 | 50±10 |
| 8 | 90±9 | 45±9 |

Survival of desiccated apices decreased in line with the bead moisture content, from 80% for 35% MC to 30% for 18% MC (Table 3). No survival was obtained after freezing apices in liquid nitrogen.

**Table 3:** Effect of bead moisture content on the survival of desiccated (-LN) and cryopreserved (+LN) pineapple (variety Puerto Rico) apices. Apices were pregrown for 5 d in medium with 0.5M sucrose before desiccation and freezing.

|  | Survival (%) | |
|---|---|---|
| Bead moisture content (%) | -LN | +LN |
| 35 | 80±8 | 0 |
| 32 | 60±10 | 0 |
| 27 | 50±5 | 0 |
| 24 | 50±1 | 0 |
| 18 | 30±4 | 0 |

*Vitrification*

When treatment with PVS2 solution was performed at 25°C, the survival of control apices was similar whatever the sucrose concentration employed in the loading solution. Survival rates ranged from 100% without exposure to PVS2 to 20-27% after 7 h of exposure to the vitrification solution (Table 4). After cryopreservation, limited survival only (6%) was achieved after loading with 0.5M sucrose followed by 5 h of exposure to PVS2, and for loading with 0.75M sucrose followed by 1 to 7 h of exposure to PVS2 (6-19%). The highest survival rate (19%) was obtained after 5 h of exposure to PVS2.

**Table 4.** Effect of sucrose concentration in the loading solution and of duration of exposure to the PVS2 vitrification solution at 25°C on the survival (%) of control (-LN) and cryopreserved (+LN) pineapple (variety Puerto Rico) apices.

|  | Sucrose concentration in loading solution (M) | | | | | |
|---|---|---|---|---|---|---|
|  | 0.3 | | 0.5 | | 0.75 | |
| Duration of exposure to PVS2 solution (hours) | -LN | +LN | -LN | +LN | -LN | +LN |
| 0 | 100 | 0 | 100 | 0 | 100 | 0 |
| 0.5 | 100 | 0 | 100 | 0 | 100 | 0 |
| 1 | 94±5 | 0 | 90±5 | 0 | 86±8 | 9±3 |
| 2 | 90±5 | 0 | 90±7 | 0 | 82±4 | 8±4 |
| 3 | 82±3 | 0 | 80±6 | 0 | 76±4 | 8±2 |
| 5 | 71±6 | 0 | 65±6 | 6.2±2 | 55±4 | 19±5 |
| 7 | 27±2 | 0 | 55±4 | 0 | 20±4 | 6±3 |

When treatment with PVS2 was performed at 0°C, higher survival rates were generally achieved than at 25°C (Table 5). Lower survival rates were obtained after extended exposure durations to PVS2 of apices loaded with higher sucrose concentrations. After cryopreservation, survival (8%) was achieved after loading with 0.5M sucrose followed by 9 h of exposure to PVS2, and after loading with 0.75M sucrose followed by 3 to 9 h of exposure to PVS2 (9-65%). The highest survival rate (65%) was obtained after 7 h of exposure to PVS2.

**Table 5.** Effect of sucrose concentration in the loading solution and of duration of exposure to the PVS2 vitrification solution at 0°C on the survival of control (-LN) and cryopreserved (+LN) pineapple (variety Puerto Rico) apices.

| Duration of exposure to PVS2 solution (hours) | Sucrose concentration in loading solution (M) | | | | | |
| --- | --- | --- | --- | --- | --- | --- |
| | 0.3 | | 0.5 | | 0.75 | |
| | -LN | +LN | -LN | +LN | -LN | +LN |
| 0 | 100 | 0 | 100 | 0 | 100 | 0 |
| 0.5 | 100 | 0 | 100 | 0 | 100 | 0 |
| 1 | 100 | 0 | 100 | 0 | 100 | 0 |
| 2 | 100 | 0 | 90±5 | 0 | 90±7 | 0 |
| 3 | 100 | 0 | 90±5 | 0 | 90±6 | 9±2 |
| 4 | 100 | 0 | 90±5 | 0 | 88±4 | 20±5 |
| 5 | 90±5 | 0 | 90±8 | 0 | 86±8 | 36±4 |
| 7 | 90±5 | 0 | 87±3 | 0 | 80±5 | 65±6 |
| 9 | 90±7 | 0 | 60±6 | 8±2 | 50±6 | 33±4 |

The optimal conditions established for freezing apices of the variety Puerto Rico proved successful in achieving survival of apices of two additional pineapple varieties, even though the survival rates measured after cryoprotection and freezing were markedly lower (Table 6). Recovery of apices always occurred directly without intermediate callus formation and numerous new plantlets could be obtained. Apices which did not survive after cryopreservation rapidly became totally black.

**Table 6.** Effect of vitrification protocol on survival of apices from 3 different pineapple varieties before (-LN) and after (+LN) cryopreservation. After loading with 1M glycerol and 0.75M sucrose for 25 min, apices were treated with PVS2 solution at 0°C for 7 h before freezing.

| | Survival (%) | |
| --- | --- | --- |
| Variety | -LN | +LN |
| Puerto Rico | 80±8 | 65±6 |
| Perolera | 50±6 | 35±7 |
| Smooth Cayenne | 50±7 | 25±4 |

## DISCUSSION/CONCLUSION

Encapsulation-dehydration did not permit successful cryopreservation of pineapple apices under the conditions experimented. These negative results can be related to the high sensitivity of pineapple apices to sucrose and dehydration. Indeed, pregrowth in media with sucrose concentrations higher than 0.5M was detrimental to survival and a prolonged treatment in 0.5M sucrose was required to improve survival after desiccation, but it was not sufficient to obtain survival of apices after freezing. The viability loss observed after freezing may be due to the crystallization of remaining freezable water upon freezing. This detrimental crystallization might be avoided by slowly freezing the encapsulated apices, which would result in removing the remaining freezable water by means of freeze-induced dehydration. Several plant materials cryopreserved by the encapsulation-dehydration technique have required a slow freezing regime to achieve optimal survival (2, 3, 11).

By contrast, survival of pineapple apices after cryopreservation was achieved using the vitrification technique, which has been successfully employed for freezing apices of a large number of different crops (5, 6, 9, 10, 13, 14, 15, 17). Optimal conditions for vitrification of pineapple apices included a 2-d preculture on semi-solid MS medium supplemented with 0.3M sucrose, loading treatment for 25 min in medium with 0.75M sucrose + 1M glycerol and dehydration at 0°C for 7 h with PVS2 before rapid immersion in liquid nitrogen.

This study emphasized the beneficial effects of applying a loading treatment to apices before dehydration with the PVS2 vitrification solution, and of performing the treatment with PVS2 at 0°C instead of room temperature. The originality of the protocol developed with pineapple apices was the extended duration (7 h) of treatment with PVS2 required to achieve optimal survival of apices after freezing, in comparison with the much shorter optimal durations (around 1-1.5 h) requested with most materials (4, 6, 7, 12). This result is probably due to the large size and compact structure of the pineapple apices employed in our experiments: the apices were around 3 mm long, and comprised the apical dome tightly covered by 2-3 leaf primordia with a very thick cuticle. Long treatment durations were needed for the vitrification solution to sufficiently dehydrate these very compact structures.

In conclusion, this study has demonstrated that vitrification can be successfully employed for freezing apices of several pineapple varieties. Various modifications of the protocol should allow improvement of the survival rates achieved and make it applicable to a larger number of varieties.

## ACKNOWLEDGEMENTS

The authors gratefully acknowledge the technical and financial support provided by IPGRI to the collaborative project between CNIC and Centro de Bioplantas-UNICA.